\documentclass[12pt, a4paper]{article}

    \usepackage{graphicx}

\begin{document}

    \title{Dark Energy
    \footnote{Invited talk at the Seventh Hungarian Relativity Workshop,
    10-15 August, 2003,
    S\'{a}rospatak, Hungary.}}

    \author{Norbert Straumann\\
        Institute for Theoretical Physics University of Zurich,\\
        CR--8057 Zurich, Switzerland}

    \maketitle

    \begin{abstract}
    After some remarks about the history and the mystery of
    the vacuum energy I shall review the current evidence for
    a cosmologically significant nearly homogeneous exotic
    energy density with negative pressure (`Dark Energy').
    Special emphasis will be put on the recent polarization
    measurements by WMAP and their implications. I shall
    conclude by addressing the question: Do the current
    astronomical observations really imply the existence of a
    dominant dark energy component?
    \end{abstract}

    \section{Introduction}

    The new results of WMAP have strengthened the evidence
    that the recent ($z<1$) Universe is dominated by an exotic
    nearly homogeneous energy density with \emph{negative}
    pressure.The simplest candidate for this so-called \emph{Dark
    Energy} is a cosmological term in Einstein's field
    equations, a possibility that has been considered during
    all the history of relativistic cosmology. Independently
    of what the nature of this energy is, one thing is clear
    since a long time:
    The energy density belonging to the cosmological
    constant is not larger than the critical cosmological
    density, and thus incredibly small by particle physics
    standards. This is a profound mystery, since we expect
    that all sorts of \emph{vacuum energies} contribute to the
    effective cosmological constant.

       is point a second puzzle has to be emphasized, because of
  which it is hard to believe that the vacuum energy
  constitutes the missing two thirds of the average energy
  density of the \emph{present} Universe. If this would be
  the case, we would also be confronted with the following
  \emph{cosmic coincidence} problem: Since the vacuum energy density is
  constant in time -- at least after the QCD phase
  transition --, while the matter energy density decreases
  as the Universe expands, it would be more than surprising
  if the two would be comparable just at about the present
  time, while their ratio was tiny in the early Universe and would
  become very large in the distant future.  The goal of so-called
  {\it quintessence models} is to avoid such an extreme fine-tuning.
  In many ways people thereby repeat what has been done in
  inflationary cosmology. The main motivation there was, as is
  well-known, to avoid excessive fine tunings of standard big bang
  cosmology (horizon and flatness problems). -- In this talk I am
  not going to say more on this topical subject. I want to
  emphasize, however, that the quintessence models do {\it not}
  solve the first problem; so far also not the second one.

  \section{History and mystery of the vacuum energy}

  Before reviewing the current evidence for a nonvanishing
  vacuum energy or some effective equivalent, it may not be
  out of place to begin with some scattered historical
  remarks. (For a more extended discussion, see \cite{N1} and
  \cite{N2}.) I begin with the classical aspect of the
  historical development.

  As is well-known, Einstein introduced the cosmological
  term when he applied general relativity the first time to
  cosmology \cite{E3}. Presumably the main reason why
  Einstein turned so soon after the completion of general
  relativity to cosmology had much to do with Machian ideas
  on the origin of inertia, which played in those years an
  important role in Einstein's thinking. His intention was
  to eliminate all vestiges of absolute space. He was, in
  particular, convinced that isolated masses cannot impose a
  structure on space at infinity. Einstein was actually
  thinking about the problem regarding the choice of boundary
  conditions at infinity already in spring 1916. In a letter
  to Michele Besso from 14 May 1916 he also mentions the possibility
  of the world being finite. A few month later he expanded on this in
  letters to Willem de Sitter. It is along these lines that
  he postulated that he postulated a Universe that is spatially
  finite and closed, a Universe in which no boundary conditions are needed. He
  then believed that this was the only way to satisfy what
  he later \cite{E4a} named \emph{Machs principle}, in the
  sense that the metric field should be determined uniquely
  by the energy-momentum tensor.

  In addition, Einstein assumed that the Universe was {\it static}. This
  was not unreasonable at the time, because the relative velocities of the
  stars as observed were small. (Recall that astronomers only
  learned later that spiral nebulae are independent star systems
  outside the Milky Way. This was definitely established when in
  1924 Hubble found that there were Cepheid variables in Andromeda
  and also in other galaxies.)

  These two assumptions were, however, not compatible with
  Einstein's original field equations. For this reason, Einstein
  added the famous $\Lambda$-term, which is compatible with the
  principles of general relativity, in particular with the
  energy-momentum law $\nabla_\nu T^{\mu\nu}=0$ for matter.

  To de Sitter Einstein emphasized in a letter on 12 March 1917, that
  his cosmological model was intended primarily to settle the question ``whether
  the basic idea of relativity can be followed through its
  completion, or whether it leads to contradictions''. And he adds
  whether the model corresponds to reality was another matter.

  Only later Einstein came to realize that Mach's
  philosophy is predicated on an antiquated ontology that seeks to
  reduce the metric field to an epiphenomenon of matter. It became
  increasingly clear to him that the metric field has an independent
  existence, and his enthusiasm for what he called Mach's principle
  later decreased. In a letter to F.Pirani he wrote in 1954: {\it
  ``As a matter of fact, one should no longer speak of Mach's
  principle at all.'' } \cite{P5}. GR still preserves
  ome remnant of Newton's absolute space and time.

Surprisingly to Einstein, de Sitter discovered in the same year, 1917, a completely
different static cosmological model which also incorporated the cosmological constant, but
was {\it anti-Machian}, because it contained no matter \cite{S6}. For this reason, Einstein
tried to discard it on various grounds (more on this below). The original form of the
metric was:

\[  g = \Bigl[ 1 - (\frac{r}{R})^2 \Bigr] dt^2 - \frac{dr^2}{1-(\frac{r}{R})^2}
-r^2(d\vartheta^2 + \sin^2\vartheta d\varphi^2 ).\] Here, the spatial part is the standard
metric of a three-sphere of radius $R$, with $R = (3/\Lambda)^{1/2}$. The model had one
very interesting property: For light sources moving along static world lines there is a
gravitational redshift, which became known as the {\it de Sitter effect}. This was thought
to have some bearing on the redshift results obtained by Slipher. Because the fundamental
(static) worldlines in this model are not geodesic, a freely- falling object released by
any static observer will be seen by him to accelerate away, generating also local velocity
(Doppler) redshifts corresponding to {\it peculiar velocities}. In the second edition of
his book \cite{E7}, published in 1924, Eddington writes about this:

{\it ``de Sitter's theory gives a double explanation for this motion of recession; first
there is a general tendency to scatter (...); second there is a general displacement of
spectral lines to the red in distant objects owing to the slowing down of atomic vibrations
(...), which would erroneously be interpreted as a motion of recession.''}

I do not want to enter into all the confusion over the de Sitter universe. One source of
this was the apparent singularity at $r=R=(3/\Lambda)^{1/2}$. This was at first thoroughly
misunderstood even by Einstein and Weyl. ( `The Einstein-de Sitter-Weyl-Klein Debate' is
now published in Vol.8 of the {\it Collected Papers} \cite{E2a}.) At the end, Einstein had
to acknowledge that de Sitter's solution is fully regular and matter-free and thus indeed a
counter example to Mach's principle. But he still discarded the solution as physically
irrelevant because it is not globally static. This is clearly expressed in a letter from
Weyl to Klein, after he had discussed the issue during a visit of Einstein in Zurich
\cite{W8}. An important discussion of the redshift of galaxies in de Sitter's model by H.
Weyl in 1923 should be mentioned. Weyl introduced an expanding version of the de Sitter
model \cite{W9}. For {\it small} distances his result reduced to what later became known as
the Hubble law \footnote{I recall that the de Sitter model has many different
interpretations, depending on the class of fundamental observers that is singled out.}.
Independently of Weyl, Cornelius Lanczos introduced in 1922 also a non-stationary
interpretation of de Sitter's solution in the form of a Friedmann spacetime with a positive
spatial curvature \cite{L10}. In a second paper he also derived the redshift for the
non-stationary interpretation \cite{L11}.

Until about 1930 almost everybody believed that the Universe was static, in spite of the
two fundamental papers by Friedmann \cite{F12} in 1922 and 1924 and Lema\^{\i}tre's
independent work \cite{L13} in 1927. These path breaking papers were in fact largely
ignored. The history of this early period has -- as is often the case -- been distorted by
some widely read documents. Einstein too accepted the idea of an expanding Universe only
much later. After the first paper of Friedmann, he published a brief note claiming an error
in Friedmann's work; when it was pointed out to him that it was his error, Einstein
published a retraction of his comment, with a sentence that luckily was deleted before
publication: {\it ``[Friedmann's paper] while mathematically correct is of no physical
significance''}. In comments to Lema\^{\i}tre during the Solvay meeting in 1927, Einstein
again rejected the expanding universe solutions as physically unacceptable. According to
Lema\^{\i}tre, Einstein was telling him: {\it ``Vos calculs sont corrects, mais votre
physique est abominable''}. On the other hand, I found in the archive of the ETH many years
ago a postcard of Einstein to Weyl from 1923, related to Weyl's reinterpretation of de
Sitter's solution, with the following interesting sentence: {\it ``If there is no
quasi-static world, then away with the cosmological term''}. This shows once more that
history is not as simple as it is often presented.

It also is not well-known that Hubble interpreted his famous results on the redshift of the
radiation emitted by distant `nebulae' in the framework of the de Sitter model, as was
suggested by Eddington.

 The general attitude is well illustrated by the following remark
of Eddington at a Royal Society meeting in January, 1930: {\it ``One puzzling question is
why there should be only two solutions. I suppose the trouble is that people look for
static solutions.''}

Lema\^{\i}tre, who had been for a short time a post-doctoral student of Eddington, read
this remark in a report to the meeting published in \emph{Observatory}, and wrote to
Eddington pointing out his 1927 paper. Eddington had seen that paper, but had completely
forgotten about it. But now he was greatly impressed and recommended Lema\^{\i}tre's work
in a letter to \emph{Nature}. He also arranged for a translation which appeared in MNRAS
\cite{L14}.

 Lema\^{\i}tre's successful explanation of Hubble's
discovery finally changed the viewpoint of the majority of workers in the field. At this
point Einstein {\it rejected the cosmological term as superfluous and no longer justified}
\cite{E15}. At the end of the paper, in which he published his new view, Einstein adds some
remarks about the age problem which was quite severe without the $\Lambda$-term, since
Hubble's value of the Hubble parameter was then about seven times too large. Einstein is,
however, not very worried and suggests two ways out. First he says that the matter
distribution is in reality inhomogeneous and that the approximate treatment may be
illusionary. Then he adds that in astronomy one should be cautious with large
extrapolations in time.

Einstein repeated his new standpoint also much later  \cite{E16}, and this was adopted by
many other influential workers, e.g., by Pauli \cite{P17}. Whether Einstein really
considered the introduction of the $\Lambda$-term as ``the biggest blunder of his life''
appears doubtful to me. In his published work and letters I never found such a strong
statement. Einstein discarded the cosmological term just for simplicity reasons. For a
minority of cosmologists (O.Heckmann, for example  \cite{H18}), this was not sufficient
reason. Paraphrasing Rabi, one might ask: `who ordered it away'?

At this point I want to leave the classical discussion of the $\Lambda$-term, but let me
add a few remarks about the quantum aspect of the $\Lambda$-problem, where it really
becomes very serious. Since quantum physicists had so many other problems, it is not
astonishing that in the early years they did not worry about this subject. An exception was
Pauli, who wondered in the early 1920s whether the zero-point energy of the radiation field
could be gravitationally effective. He estimated the influence of the zero-point energy of
the radiation field -- cut off at the classical electron radius -- on the radius of the
universe, and came to the conclusion that it {\it ``could not even reach to the moon''}.
(For more on this, see \cite{N2}. Pauli's only published remark on his considerations can
be found in his Handbuch article on quantum mechanics \cite{P19}, in the section on the
quantization of the radiation field, where he says: {\it `Also, as is obvious from
experience, the [zero-point energy] does not produce any gravitational field.'})

For decades nobody else seems to have worried about contributions of quantum fluctuations
to the cosmological constant, although physicists learned after Dirac's hole theory that
the vacuum state in quantum field theory is not an empty medium, but has interesting
physical properties. As far as I know, the first who came back to possible contributions of
the vacuum energy density to the cosmological constant was Zel'dovich. He discussed this
issue in two papers \cite{Z20} during the third renaissance period of the $\Lambda$-term,
but before the advent of spontaneously broken gauge theories. The following remark by him
is particularly interesting. Even if one assumes completely ad hoc that the zero-point
contributions to the vacuum energy density are exactly cancelled by a bare term, there
still remain higher-order effects. In particular, {\it gravitational} interactions between
the particles in the vacuum fluctuations are expected on dimensional grounds to lead to a
gravitational self-energy density of order $G\mu^6$, where $\mu$ is some cut-off scale.
Even for $\mu$ as low as 1 GeV (for no good reason) this is about 9 orders of magnitude
larger than the observational bound.

This illustrates that there is something profound that we do not understand at all,
certainly not in quantum field theory ( so far also not in string theory).  We are unable
to calculate the vacuum energy density in quantum field theories, like the Standard Model
of particle physics. But we can attempt to make what appear to be reasonable
order-of-magnitude estimates for the various contributions. All expectations  are
\textbf{in gigantic conflict with the facts} (see, e.g., [1]). Trying  to arrange the
cosmological constant to be zero is unnatural in a technical sense. It is like enforcing a
particle to be massless, by fine-tuning the parameters of the theory when there is no
symmetry principle which implies a vanishing mass. The vacuum energy density is unprotected
from large quantum corrections. This problem is particularly severe in field theories with
spontaneous symmetry breaking. In such models there are usually several possible vacuum
states with different energy densities. Furthermore, the energy density is determined by
what is called the effective potential, and this is a {\it dynamical} object. Nobody can
see any reason why the vacuum of the Standard Model we ended up as the Universe cooled, has
-- for particle physics standards --  an almost vanishing energy density. Most probably, we
will only have a satisfactory answer once we shall have a theory which successfully
combines the concepts and laws of general relativity about gravity and spacetime structure
with those of quantum theory.

\section{Microwave background anisotropies}

Investigations of the cosmic microwave background have presumably contributed most to the
remarkable progress in cosmology during recent years (For a recent review, see \cite{H21}.
Beside its spectrum, which is Planckian to an incredible degree, we also can study the
temperature fluctuations over the ``cosmic photosphere'' at a redshift $z\approx1100$.
Through these we get access to crucial cosmological information (primordial density
spectrum, cosmological parameters, etc). A major reason for why this is possible relies on
the fortunate circumstance that the fluctuations are tiny ($\sim 10^{-5}$ ) at the time of
recombination. This allows us to treat the deviations from homogeneity and isotropy for an
extended period of time perturbatively, i.e., by linearizing the Einstein and matter
equations about solutions of the idealized Friedmann-Lema\^{\i}tre models. Since the
physics is effectively {\it linear}, we can accurately work out the {\it evolution} of the
perturbations during the early phases of the Universe, given a set of cosmological
parameters. Confronting this with observations, tells us a lot about the cosmological
parameters as well as the initial conditions, and thus about the physics of the very early
Universe. Through this window to the earliest phases of cosmic evolution we can, for
instance, test general ideas and specific models of inflation.

\subsection{Qualitative remarks}

Let me begin with some qualitative remarks, before I go into more technical details. Long
before recombination (at temperatures $T>6000 K$, say) photons, electrons and baryons were
so strongly coupled that these components may be treated together as a single fluid. In
addition to this there is also a dark matter component. For all practical purposes the two
interact only gravitationally. The investigation of such a two-component fluid for small
deviations from an idealized Friedmann behavior is a well-studied application of
cosmological perturbation theory.

At a later stage, when decoupling is approached, this approximate treatment breaks down
because the mean free path of the photons becomes longer (and finally `infinite' after
recombination). While the electrons and baryons can still be treated as a single fluid, the
photons and their coupling to the electrons have to be described by the general
relativistic Boltzmann equation. The latter is, of course, again linearized about the
idealized Friedmann solution. Together with the linearized fluid equations (for baryons and
cold dark matter, say), and the linearized Einstein equations one arrives at a complete
system of equations for the various perturbation amplitudes of the metric and matter
variables. There exist widely used codes e.g. CMBFAST \cite{S22}, that provide the CMB
anisotropies -- for given initial conditions -- to a precision of about 1\%. A lot of
qualitative and semi-quantitative insight into the relevant physics can, however, be gained
by looking at various approximations of the basic dynamical system.

Let us first discuss the temperature fluctuations. What is observed is the temperature
autocorrelation:
\begin{equation}
 C(\vartheta ):= \langle \frac{\Delta T(\mathbf{n})}{T}\cdot
\frac{\Delta T(\mathbf{n'})}{T}\rangle\\
= \sum_{l=2}^\infty \frac{2l+1}{4\pi} C_l P_l(\cos \vartheta) ,
\end{equation}
where $\vartheta$ is the angle between the two directions of observation $\mathbf{n},
\mathbf{n'}$, and the average is taken ideally over all sky. The {\it angular power
spectrum} is by definition $\frac{l(l+1)}{2\pi}C_l \; \; versus \; \;l \; \; (\vartheta
\simeq \pi /l ).$

A characteristic scale, which is reflected in the observed CMB anisotropies, is the sound
horizon at last scattering, i.e., the distance over which a pressure wave can propagate
until decoupling. This can be computed within the unperturbed model and subtends about half
a  degree on the sky for typical cosmological parameters. For scales larger than this sound
horizon the fluctuations have been laid down in the very early Universe. These have been
detected by the COBE satellite. The (gauge invariant brightness) temperature perturbation
$\Theta = \Delta T/T$ is dominated by the combination of the intrinsic temperature
fluctuations and gravitational redshift or blueshift effects. For example, photons that
have to climb out of potential wells for high-density regions are redshifted. One can show
that these effects combine for adiabatic initial conditions to $\frac{1}{3}\Psi$, where
$\Psi$ is one of the two gravitational Bardeen potentials. The latter, in turn, is directly
related to the density perturbations. For scale-free initial perturbations and almost
vanishing spatial curvature the corresponding angular power spectrum of the temperature
fluctuations turns out to be nearly flat (Sachs-Wolfe plateau; see, e.g., Fig.3 of
Ref.[1]).

On the other hand, inside the sound horizon before decoupling, acoustic, Doppler,
gravitational redshift, and photon diffusion effects combine to the spectrum of small angle
anisotropies. These result from gravitationally driven synchronized acoustic oscillations
of the photon-baryon fluid, which are damped by photon diffusion (for details, see again
[1]).

A particular realization of $\Theta(\mathbf{n})$, such as the one accessible to us (all sky
map from our location), cannot be predicted. Theoretically, $\Theta $ is a random field
$\Theta(\mathbf{x},\eta,\mathbf{n})$, depending on the conformal time $\eta$, the spatial
coordinates, and the observing direction $\mathbf{n}$. Its correlation functions should be
rotationally invariant in $\mathbf{n}$, and respect the symmetries of the background time
slices. If we expand $\Theta$ in terms of spherical harmonics,
\begin{equation}
\Theta(\mathbf{n}) = \sum_{lm} a_{lm} Y_{lm}(\mathbf{n}),
\end{equation}
the random variables $a_{lm}$ have to satisfy
\begin{equation}
\langle a_{lm}\rangle = 0,\;\;\;  \langle a_{lm}^\star a_{l'm'}\rangle =
\delta_{ll'}\delta_{mm'}C_l(\eta),
\end{equation}
where the $C_l(\eta)$ depend only on $\eta$. Hence the correlation function at the present
time $\eta_0$ is given by (1), where $C_l = C_l(\eta_0)$, and the bracket now denotes the
statistical average. Thus,
\begin{equation}
C_l = \frac{1}{2l+1}\langle\sum_{m=-l}^la_{lm}^\star a_{lm}\rangle.
\end{equation}
The standard deviations $\sigma(C_l)$ measure a fundamental uncertainty in the knowledge we
can get about the $C_l$'s. These are called {\it cosmic variances}, and are most pronounced
for low $l$. In simple inflationary models the $a_{lm}$ are Gaussian distributed, hence
\begin{equation}
\frac{ \sigma(C_l)}{C_l} = \sqrt{\frac{2}{2l+1}}.
\end{equation}
Therefore, the limitation imposed on us (only one sky in one universe) is small for large
$l$.

\subsection{Boltzmann hierarchy}

The brightness temperature fluctuation can be obtained from the perturbation of the photon
distribution function by integrating over the magnitude of the photon momenta. The
linearized Botzmann equation can then be translated into an equation for $\Theta$, which we
now regard as a function of $\eta, x^i$, and $\gamma^j$, where the $\gamma^j$ are the
directional cosines of the momentum vector relative to an orthonormal triad field of the
unperturbed spatial metric with curvature $K$. Next one performs a harmonic decomposition
of $\Theta$, which reads for the spatially flat case ($K=0$)
\begin{equation}
\Theta(\eta,\mbox{\boldmath$x$},\mbox{\boldmath$\gamma$}) = (2\pi)^{-3} \int d^3k\sum_l
\theta_l(\eta,k) G_l(\mbox{\boldmath$x$},\mbox{\boldmath$\gamma$};\mbox{\boldmath$k$}),
\end{equation}
where
\begin{equation}
G_l(\mbox{\boldmath$x$},\mbox{\boldmath$\gamma$};\mbox{\boldmath$k$}) = (-i)^l
P_l(\mbox{\boldmath$\hat{k}\cdot \gamma$}) \exp(i\mbox{\boldmath$k\cdot x$}).
\end{equation}
The dynamical variables $\theta_l(\eta)$ are the {\it brightness moments}, and should be
regarded as random variables. Boltzmann's equation implies the following hierarchy of
ordinary differential equations for the brightness moments\footnote{In the literature the
normalization of the $\theta_l$ is sometimes chosen differently: $\theta_l\rightarrow
(2l+1)\theta_l$.} $\theta_l(\eta)$ (if polarization effects are neglected):
\begin{eqnarray}
\theta_0' & = &-\frac{1}{3}k \theta_1 - \Phi',  \\
\theta_1' & = & k\Bigl(\theta_0 + \Psi -\frac{2}{5}\theta_2 \Bigr)
-\dot{\tau}(\theta_1 -V_b),  \\
\theta_2' & = & k\Bigl(\frac{2}{3}\theta_1 -\frac{3}{7}\theta_3\Bigr) -
\dot{\tau} \frac{9}{10}\theta_2, \\
\theta_l' & = & k\Bigl(\frac{l}{2l-1}\theta_{l-1} -\frac{l+1}{2l+3}\theta_{l+1}\Bigr),  \;
\; \; l>2.
\end{eqnarray}
Here, $V_b$ is the gauge invariant scalar velocity perturbation of the baryons,
$\dot{\tau}=x_e n_e \sigma_T a/a_0$, where $a$ is the scale factor, $x_e n_e $ the
unperturbed free electron density ($x_e$ =  ionization fraction), and $\sigma_T$ the
Thomson cross section. Moreover, $\Phi$ and $\Psi$ denote the Bardeen potentials. (For
further details, see Sect.6 of [1].)

The $C_l$ are determined by an integral over $k$, involving a primordial power spectrum (of
curvature perturbations) and the $|\theta_l(\eta)|^2$, for the corresponding initial
conditions (their transfer functions).

This system of equations is completed by the linearized fluid and Einstein equations.
Various approximations for the Boltzmann hierarchy provide already a lot of insight. In
particular, one can very nicely understand how damped acoustic oscillations are generated,
and in which way they are influenced by the baryon fraction (again, see Ref.[1]). A typical
theoretical CMB spectrum is shown in Fig.3 of [1]. (Beside the scalar contribution in the
sense of cosmological perturbation theory, considered so far, the tensor contribution due
to gravity waves is also shown there.)

\section{Polarization}

A polarization map of the CMB radiation provides important additional information to that
obtainable from the temperature anisotropies. For example, we can get constraints about the
epoch of reionization. Most importantly, future polarization observations may reveal a
stochastic background of gravity waves, generated in the very early Universe. In this
section we give a brief introduction to the study of CMB polarization.

The mechanism which partially polarizes the CMB radiation is similar to that for the
scattered light from the sky. Consider first scattering at a single electron of unpolarized
radiation coming in from all directions . Due to the familiar polarization dependence of
the differential Thomson cross section, the scattered radiation is, in general, polarized.
It is easy to compute the corresponding Stokes parameters. Not surprisingly, they are not
all equal to zero if and only if the intensity distribution of the incoming radiation has a
non-vanishing quadrupole moment. The Stokes parameters $Q$ and $U$ are proportional  to the
overlap integral with the combinations $Y_{2,2} \pm Y_{2,-2}$ of the spherical harmonics,
while $V$ vanishes.) This is basically the reason why a CMB  polarization map traces (in
the tight coupling limit) the quadrupole temperature distribution on the last scattering
surface.

The polarization tensor of an all sky map of the CMB radiation can be parametrized in
temperature fluctuation units, relative to the orthonormal basis $\{d\vartheta,
\sin\vartheta\; d\varphi\}$ of the two sphere, in terms of the Pauli matrices as
$\Theta\cdot 1 + Q\sigma_3 + U\sigma_1 + V\sigma_2$. The Stokes parameter $V$ vanishes (no
circular polarization). Therefore, the polarization properties can be described by the
following symmetric trace-free tensor on $S^2$:
\begin{equation}
(\mathcal{P}_{ab}) = \left(\begin{array}{cc}
Q&U\\
U&-Q\\
\end{array}\right).
\end{equation}

As for gravity waves, the components $Q$ and $U$ transform under a rotation of the 2-bein
by an angle $\alpha$ as
\begin{equation}
Q \pm iU \rightarrow e^{\pm 2i\alpha}(Q\pm iU),
\end{equation}
and are thus of spin-weight 2. $\mathcal{P}_{ab}$ can be decomposed uniquely into
\emph{`electric'} and \emph{`magnetic'} parts:
\begin{equation}
\mathcal{P}_{ab} = E_{;ab} - \frac{1}{2}g_{ab} \Delta E + \frac{1}{2}(\varepsilon_a{}^c
B_{;bc} +\varepsilon_b{}^c B_{;ac}).
\end{equation}
Expanding here the scalar functions $E$ and $B$ in terms of spherical harmonics, we obtain
an expansion of the form
\begin{equation}
\mathcal{P}_{ab} = \sum_{l=2}^{\infty} \sum_{m} \left[a^E_{(lm)}Y^E_{(lm)ab} +
a^B_{(lm)}Y^B_{(lm)ab} \right]
\end{equation}
in terms of the tensor harmonics:
\begin{equation}
Y^E_{(lm)ab}: = N_l(Y_{(lm);ab} -\frac{1}{2}g_{ab}Y_{(lm);c}{}^c),\;\; Y^B_{(lm)ab}: =
\frac{1}{2}N_l(Y_{(lm);ac} \varepsilon^c{}_b + a\leftrightarrow b),
\end{equation}
where $l\geq 2$ and
\[ N_l \equiv \left(\frac{2(l-2)!}{(l+2)!}\right)^{1/2}.\]
Equivalently, one can write this as
\begin{equation}
Q+iU = \sqrt{2}\sum_{l=2}^\infty\sum_{m}\left[a^E_{(lm)}+ ia^B_{(lm)}\right] \,_{2\!}Y_l^m,
\end{equation}
where $\,_{s\!}Y_l^m$ are the spin-s harmonics.

As in Eq.(2) the multipole moments $a^E_{(lm)}$ and $a^B_{(lm)}$ are random variables, and
we have equations analogous to (3):
\begin{equation}
C_l^{TE} = \frac{1}{2l+1}\sum_{m} \;\langle a_{lm}^{\Theta \star} a_{lm}^E \rangle ,\;\;
etc.
\end{equation}
(We have now put the superscript $\Theta$ on the $a_{lm}$ of the temperature fluctuations.)
The $C_l$'s determine the various angular correlation functions. For example, one easily
finds
\begin{equation}
\langle\Theta(\mbox{\boldmath$n$})Q(\mbox{\boldmath$n'$})\rangle = \sum_{l}\;
C_l^{TE}\frac{2l+1}{4\pi}N_l P_l^2(\cos \vartheta).
\end{equation}

For the space-time dependent Stokes parameters $Q$ and $U$ of the radiation field we can
perform a normal mode decomposition analogous to (6). If, for simplicity, we again consider
only scalar perturbations this reads
\begin{equation}
Q\pm iU = (2\pi)^{-3} \int d^3k\sum_l(E_l \pm iB_l)\,_{\pm 2\!}G^0_l,
\end{equation}
where
\begin{equation}
\,_{s\!}G_l^m(\mbox{\boldmath$x$},\mbox{\boldmath$\gamma$};\mbox{\boldmath$k$}) =
(-i)^l\left(\frac{2l+1}{4\pi}\right)^{1/2} \,_{s\!}Y_l^m(\mbox{\boldmath$ \gamma$})
\exp(i\mbox{\boldmath$k\cdot x$}),
\end{equation}
if the mode vector $\mathbf{k}$ is chosen as the polar axis. (Note that $G_l$ in (7) is
equal to $\,_{0\!}G_l^0$.)

The Boltzmann equation implies a coupled hierarchy for the moments $\theta_l, E_l$, and
$B_l$ \cite{H23}, \cite{H24}. It turns out that the $B_l$ vanish for scalar perturbations.
Non-vanishing magnetic multipoles would be a unique signature for a spectrum of gravity
waves. In a sudden decoupling approximation, the present electric multipole moments can be
expressed in terms of the brightness quadrupole moment on the last scattering surface and
spherical Bessel functions as
\begin{equation}
\frac{E_l(\eta_0,k)}{2l+1}\simeq \frac{3}{8}\theta_2(\eta_{dec},k)
\frac{l^2j_l(k\eta_0)}{(k\eta_o)^2}.
\end{equation}
Here one sees how the observable $E_l$'s trace the quadrupole temperature anisotropy on the
last scattering surface. In the tight coupling approximation the latter is proportional to
the dipole moment $\theta_1$.

\section{Observational results}

In recent years several experiments gave clear evidence for multiple peaks in the angular
temperature power spectrum at positions expected on the basis of the simplest inflationary
models and big bang nucleosynthesis \cite{S25}. These results have been confirmed and
substantially improved by WMAP \cite{B26} (see, in particular, Fig.12 of Ref.[26]).

In spite of the high accuracy of the data, it is not possible to extract unambiguously
cosmological parameters, because there are intrinsic degeneracies, especially when tensor
modes are included. These can only be lifted if other cosmological information is used.
Beside the supernova results, use has been made for instance of the available information
for the galaxy power spectrum (in particular from the 2-degree-Field Galaxy Redshift Survey
(2dFGRS)), and limits for the Hubble parameter. For example, if one adds to the CMB data
the well-founded constraint $H_0\geq 50\;km/s/Mpc$, then the total density parameter
$\Omega_{tot}$ has to be in the range $0.98 < \Omega_{tot} < 1.08$ (95 \%). The Universe is
thus \emph{spatially almost flat}. In what follows we therefore always assume $K=0$.

Table 1 is extracted from the extended analysis \cite{S27} of the WMAP data and other
cosmological information. It shows the 68\% confidence ranges for some of the cosmological
parameters for two types of fits, assuming a $\Lambda$CDM model. In the first only the CMB
data are used (but tensor modes are included), while in the second these data are combined
with the 2dFGRS power spectrum (assuming adiabatic, Gaussian initial conditions described
by power laws). \vspace{1cm}

\begin{tabular}{|l||c|r|}
\multicolumn{3}{c}{\textbf{Table 1.}} \\  \hline Parameter & CMB alone & CMB and 2dFGRS\\
\hline\hline
$\Omega_b h_0^2$ & 0.024 $\pm$ 0.001 & 0.023 $\pm$ 0.001\\
$\Omega_M h_0^2$ & 0.14 $\pm$ 0.02   & 0.134 $\pm$ 0.006   \\
$h_0 $ &   0.72 $\pm$ 0.05    & 0.71 $\pm$ 0.04  \\
$\Omega_b$  & 0.047 $\pm$ 0.006       & $\simeq$ same   \\
$\Omega_M$  & 0.29 $\pm$0.07         &  $\simeq$ same   \\
\hline
\end{tabular}

\vspace{1cm}
 Note that there is little difference between the two columns. The
 age of the Universe for these parameters is close to 14 Gyr.
 Another interesting result coming from the rise of the
 temperature-polarization correlation function at large scales
 (small $l$) is that reionization of the Universe has set in
 surprisingly early --, at a redshift of $z_r = 17\pm 5$, with a
 corresponding optical depth $\tau = 0.17\pm 0.06$.

 Before the new results possible admixtures of isocurvature modes were
 not strongly constraint. But now the measured temperature-polarization
 correlations imply that the primordial fluctuations were primarily
 \textit{adiabatic}. Admixtures of isocurvature modes do not improve the fit.

 One worry is that the quadrupole amplitude ($C_2$) measured by
 WMAP is lower than expected according to the best fit $\Lambda$CDM
 model [28]. This issue has led to lots of discussions. A recent
 reanalysis \cite{E28} of the effects of Galactic cuts indicates
 that this discrepancy is not particularly significant, being in
 the region of a few percent.

 \section{Concluding remarks}

 A wide range of astronomical data support the following
 `concordance' $\Lambda$CDM model: The Universe is spatially flat
 and dominated by vacuum energy density and weakly interacting
 cold dark matter. Furthermore, the primordial fluctuations are
 adiabatic and nearly scale invariant, as predicted in simple
 inflationary models.

 A vacuum energy with density parameter $\Omega_\Lambda \simeq 0.7$ is so
 surprising that it should be examined whether this conclusion is
 really unavoidable. Since we do not have a tested theory predicting
 the spectrum of primordial fluctuations, it appears reasonable to
 consider a wider range of possibilities than simple power laws.
 An instructive attempt in this direction has been made in
 \cite{B29}, by constructing an Einstein-de Sitter model with
 $\Omega_\Lambda = 0$, fitting the CMB data as well as the power
 spectrum of 2dFGRS. In this the Hubble constant is, however,
 required to be rather low: $H_0 \simeq 46\; km/s/Mpc$. The
 authors argue that this cannot definitely be excluded,
 because `physical' methods lead mostly to relatively low values
 of $H_0$. In order to be consistent with matter fluctuations on cluster
 scales they add relic neutrionos with degenerate masses of order eV and
 a small contribution of quintessence with zero pressure ($w=0$).
 In addition, they have to ignore the direct evidence for
 an accelerating Universe from the Hubble-diagram for distant Type
 Ia supernovae, on the basis of remaining systematic
 uncertainties.

 It is very likely that the present concordance model will
 survive, but for the time being it is healthy to remain sceptical
 until further evidence is accumulating.

\end{document}